\documentclass{article}

\usepackage{arxiv}

\usepackage[utf8]{inputenc} 
\usepackage[ngerman]{babel}
\usepackage{hyperref}       
\usepackage{url}            
\usepackage{booktabs}       
\usepackage{amsfonts}       
\usepackage{nicefrac}       
\usepackage{microtype}      
\usepackage{lipsum}

\usepackage{amsthm}
\usepackage{graphicx}
\usepackage{color}
\usepackage{xcolor,soul}
\usepackage{grffile}

\usepackage{amssymb}
\usepackage{amsmath}

\newtheorem{definition}{Definition}[section]

\newcommand{\N}{\mathbb{N}}

\newcommand{\iN}{\in \N}

\title{Bemerkungen zum paarweisen Vergleich}

\author{Stefan Lörcks}

\begin{document}
\maketitle

\begin{abstract}
Der einfache paarweise Vergleich ist ein Verfahren verschiedene Kriterien mit einer Gewichtung zu versehen. Wir zeigen, dass die Werte dieser Gewichte (insbesondere auch der maximale Wert) ausschließlich von der Anzahl der Kriterien abhängt. Darüber hinaus wird gezeigt, dass der Abstand der Gewichtungen stets gleich ist.
\end{abstract}


\section{Einleitung}
In der Nutzwertanalyse nach Zangemeister \cite{zangemeister1976nutzwertanalyse} werden verschiedene Alternativen nach ihrem Nutzwert verglichen. Dabei werden die Alternativen in verschiedenen Kriterien beurteilt. Der paarweise Vergleich ist ein Verfahren um den verschiedenen Kriterien eine Gewichtung zuzuteilen.

\section{Paarweiser Vergleich}
Wir diskutieren hier den 'einfachen' paarweisen Vergleich, wie er beispielsweise in \cite{von2018investitionstheorie} vorgestellt wird. Ein anderer Paarvergleich wird im Rahmen des Analytic Hierarchy Proces (AHP) \cite{SAATY1977234} vorgeschlagen. Wie die nachfolgende Definition zeigt,  weichen wir von dem in \cite{von2018investitionstheorie} vorgeschlagenen Verfahren insofern ab, dass wir keine Zirkelschlüsse zulassen.

\begin{definition}[Einfacher Paarweiser Vergleich ]
Gegeben sei eine Mengevon $k\iN$ Kriterien $\{K_1, K_2, \ldots,K_k\}$. Dabei wird vorausgesetzt, dass sich je zwei Kriterien paarweise vergleichen lassen, in dem Sinne, dass eine eindeutige Zuordnung möglich sei, welches Kriterium wichtiger ist. Ist beispielsweise $K_1$ wichtiger als $K_2$, so so notieren wir dies mit $K_1 \succ K_2$. Dazu gebe es Koeffizienten $a_{i,j} \in \{0,1\}, i,j \in \{1,2,\ldots,k\},i\neq j$. Dabei wird $a_{i,j}=1$ gesetzt, falls $K_i \succ K_2$ und sonst $a_{i,j}=0$.

Dabei muss stets $a_{i,j}\neq a_{j,i}$ (beziehungsweise $a_{i,j} = 1- a_{j,i}$) gelten, denn zwei Kriterien können nicht jeweils wichtiger als das Andere sein. Verallgemeinert sind bei der Festlegung der Koeffizienten Zirkelschlüsse zu vermeiden. Beispielsweise kann nicht $K_1 \succ K_2 \succ K_3 \succ K_1$ gelten. Darauf ist bei der Festlegung der Koeffizienten $a_{i,j}$ zu achten.

Die Koeffizienten $a_{i,j}$ können als Einträge einer $k \times k$ Matrix $A$ betrachtet werden, wobei die Elemente der Hauptdiagonalen unbesetzt bleiben. In dieser Anschauung entspricht die Anzahl der Einsen in der $i$-ten Zeile der Matrix der Anzahl der Kriterien, die unwichtiger sind als $K_i$. Diese Zahl ist also proportional zu der Wichtigkeit von $K_i$. Entsprechend ergibt sich die Gewichtung $w_i$ des Kriteriums $K_i$ als Normierung dieser Anzahl:
\[w_i := \dfrac{\sum \limits_{\underset{j\neq i}{j=1}}^k a_{i,j}}{\sum \limits_{i=1}^k \sum \limits_{\underset{j\neq i}{j=1}}^k a_{i,j}}\]
\end{definition}

Der nachfolgende Satz liefert die zentrale Aussage.

\textbf{(Satz)} Beim einfachen paarweisen Vergleich mit $k$ Faktoren ist die höchste mögliche Bewertung durch $\frac{2}{k}$ gegeben. Allgemein ergeben sich die vorhandenen Gewichtungen für $i \in \{0, 1, \ldots, k-1\}$ zu 
\[\frac{2i}{(k-1)k}, \]
womit die Abstände zwischen den Gewichtungen stets durch
\[\frac{2}{(k-1)k} \]
gegeben sind.\\

\emph{Bew.: } Bei $k$ Faktoren ist die Matrix des paarweisen Vergleichs von der Größe $k \times k$, wobei beim einfach paarweisen Vergleich auf der Hauptdiagonalen keine Einträge zugelassen werden. Zunächst bestimmen wir die Gesamtzahl der Einsen in der Matrix. O. B. d. A. nehmen wir an, dass alle Einsen oberhalb der Hauptdiagonalen eingetragen sind. (Dies ist zulässig, da sich die Anzahl der Einsen nicht ändert, egal ob sie ober- oder unterhalb der Diagonalen stehen.) Dann stehen in der ersten Zeile $(k-1)$ Einsen, in der zweiten Zeile $(k-2)$, \ldots, in der vorletzten Zeile eine Eins und in der letzten Zeile keine. Die Gesamtanzahl an Einsen ergibt sich demnach zu
\[(k-1) + (k-2) + \ldots +1 \overset{\text{Gauß}}{=} \frac{(k-1)k}{2}\]
In einer Zeile können überall außer auf der Hauptdiagonalen Einsen stehen. Also maximal $k-1$ Stück. Also ergibt sich die höchstmögliche Bewertung eines Faktors zu
\[\frac{k-1}{\frac{(k-1)k}{2}} = \frac{2}{k}. \]
Analog ergeben sich für die zweite Zeile $i=k-2$ Einsen usw. In der letzten Zeile sind es $i=0$ Einsen rechts der Hauptdiagonalen. Somit sind also in der Tat die möglichen Faktoren durch
\[\frac{2i}{(k-1)k}, \quad i=0,1\ldots,k-1 \]
gegeben und für $i=1,\ldots,k-1$ ergibt sich der Abstand zwischen zwei benachbarten Gewichtungen zu
\[\frac{i}{\frac{(k-1)k}{2}}-\frac{i-1}{\frac{(k-1)k}{2}} = \frac{2}{(k-1)k}. \] \hfill $\Box$

\section{Zusammenfassung}
Es wurden gezeigt, dass die maximale Gewichtung und die Abstände der Gewichtung beim einfach paarweisen Vergleich nur von der Anzahl der betrachteten Kriterien abhängt.

\bibliographystyle{unsrt}  
\bibliography{references}  

\begin{thebibliography}{1}

\bibitem{zangemeister1976nutzwertanalyse}
C.~Zangemeister.
\newblock {\em Nutzwertanalyse in der Systemtechnik: eine Methodik zur
  multidimensionalen Bewertung und Auswahl von Projektalternativen}.
\newblock Wittemann, 1976.

\bibitem{von2018investitionstheorie}
W.~Busse~von Colbe and F.~Witte.
\newblock {\em Investitionstheorie und Investitionsrechnung}.
\newblock Springer Berlin Heidelberg, 2018.

\bibitem{SAATY1977234}
T.~L. Saaty.
\newblock A scaling method for priorities in hierarchical structures.
\newblock {\em Journal of Mathematical Psychology}, 15(3):234 -- 281, 1977.

\end{thebibliography}


\end{document}